\documentstyle[12pt]{article}

\newcommand{\nc}{\newcommand}
\def\frac#1#2{{\textstyle {#1 \over #2}}}

\nc{\beq}{\begin{equation}}
\nc{\eeq}{\end{equation}}
\nc{\beqa}{\begin{eqnarray}}
\nc{\eeqa}{\end{eqnarray}}
\nc{\lsim}{\begin{array}{c}\,\sim\vspace{-21pt}\\< \end{array}}
\nc{\gsim}{\begin{array}{c}\sim\vspace{-21pt}\\> \end{array}}

\def\&{and}

\def\DS {D\!\!\!\!/}

\def\Dslash{\not{\hbox{\kern-4pt $D$}}}

\def\nc#1#2#3{           {\it Nuovo Cim.  }{\bf #1}, #2 (19#3)}
\def\np#1#2#3{           {\it Nucl. Phys. }{\bf #1}, #2 (19#3)}
\def\pl#1#2#3{           {\it Phys. Lett. }{\bf #1}, #2 (19#3)}
\def\pr#1#2#3{           {\it Phys. Rev. }{\bf #1}, #2 (19#3)}
\def\prep#1#2#3{         {\it Phys. Rep. }{\bf #1}, #2 (19#3)}
\def\prl#1#2#3{          {\it Phys. Rev. Lett. }{\bf #1}, #2 (19#3)}

\begin{document}

\title{\large{\bf
Gaugino Determinant in Supersymmetric Yang-Mills Theory}}

\author{
Stephen D.H.~Hsu\thanks{Present address: Institute for Theoretical Science,
University of Oregon, Eugene OR, 97403-5203. Email: hsu@duende.uoregon.edu}
\\ \\ Department of Physics \\ Yale University
\\ New Haven, CT 06520-8120 }

\date{April, 1997}

\maketitle

\begin{picture}(0,0)(0,0)
\put(350,315){YCTP-P4-97}
\end{picture}
\vspace{-24pt}

\begin{abstract}
We resolve an ambiguity in the sign of the determinant of a 
single Weyl fermion, such as the gaugino in supersymmetric models.
Positivity of this determinant is necessary
for application of QCD inequalities and lattice Monte Carlo methods to 
supersymmetric Yang-Mills models.

\end{abstract}

\newpage


In this note we address an ambiguity in the sign of the gaugino
determinant. This ambiguity is important, in that it affects the possible
non-perturbative methods available to study supersymmetric
Yang-Mills \cite{SYM} (SYM) theory. 
In particular, the application of QCD inequalities \cite{Peskin} as well as 
lattice Monte Carlo methods \cite{CV,Montvay} to SYM
rely on the positivity of the fermion
measure.
Because SYM is vector like (one can write a gauge invariant majorana
mass for the gaugino), one is tempted to conclude immediately that
the measure is positive definite.
However, 
the definition of the gaugino determinant is somewhat subtle, 
and some technical machinery (an index theorem in five dimensions) is necessary
to resolve the sign ambiguity.

The determinant for a single Weyl fermion cannot be straightforwardly 
defined because the Weyl operator maps the vector
space of left-handed spinors into the space of right-handed spinors, 
and therefore fails to define an eigenvalue problem \cite{AW}.
Rather, one usually defines the determinant in terms of the
eigenvalues of the Dirac operator \cite{Witten}. Naively, one simply
writes
\beq
{\rm det}~  i \Dslash_{\rm \, Weyl}    ~=~ 
( {\rm det}~  i \Dslash_{\rm \, Dirac} )^{1/2} ~.
\eeq
However, this definition leads to a sign ambiguity, as first noticed by
Witten \cite{Witten}. Suppose we define the Weyl determinant
for some fiducial background gauge configuration (which
we take here to be $A^0_{\mu} (x) = 0$)
as the product of only the {\it positive} eigenvalues of 
$i \Dslash_{\rm \, Dirac}$. Once this choice is made, there is no additional
freedom,
and the Weyl determinant is defined for all $A_{\mu} (x)$ by the condition that
it vary smoothly as the gauge field is varied (see \cite{chiralme} for further
discussion). This condition requires that we continue to take the Weyl
determinant as the product of the {\it same} eigenvalues, which flow
continuously as the gauge field is varied. The sign ambiguity arises
because some of the originally positive 
eigenvalues can become negative for some background fields.
If an odd number do so, the determinant becomes negative and therefore
spoils the positivity property of the measure. This is problematic for
Monte Carlo simulations, because the functional integral loses
its statistical interpretation.

Witten investigated these possible sign changes in the case
of $SU(2)$ with a Weyl fermion in the fundamental representation.
There, the model is intrinsically inconsistent, because the Weyl determinant
changes sign under certain toplogically non-trivial gauge transformations.
In SYM we are interested in a related behavior which, while not rendering
the theory inconsistent, would make it more 
difficult to study at the non-perturbative level.

Fortunately, one can show that for a Weyl fermion in 
the adjoint representation,
the eigenvalue flow always involves an even number of eigenvalues
crossing zero. 
Hence the sign of the gaugino determinant is constant
and can be chosen to be positive. 
We use the machinery of  
reference \cite{Witten}. 
There, it is demonstrated
that the flow of eigenvalues of the four dimensional Dirac operator can be
related to the number of zero modes of the five dimensional Dirac operator
$\DS_5$
on a cylinder $S^4 \times R$, consisting of the 
smooth interpolation of the fiducial gauge field to
the gauge field of interest, $A_\mu (x)$.
The mod two Atiyah-Singer index theorem \cite{AS}
gives the number of zero modes of $\DS_5$ modulo 2 as
twice $C(R)$ (the Casimir of the fermion representation) times
an integer topological invariant related to $\pi_4 ( G )$, where $G$
is the gauge group. In \cite{Witten} 
the index theorem is applied to cases in which
the (four dimensional) gauge field of interest is a gauge 
transform of the fiducial gauge field ($A^0_{\mu} (x) = 0$):
\beq
\label{pg}
A_\mu (x) ~=~ i U^{\dagger} \partial_{\mu} U (x)~~~.
\eeq

We do not wish to restrict ourselves to this case, as we need
the sign of the Weyl determinant for arbitrary
$A_\mu (x)$. In order to consider arbitrary gauge fields, we 
will exploit the fact that
the number of zero modes of $\DS_5$ is conserved mod 2 under
any smooth deformation of the five dimensional gauge configuration. 
This result is easy to see since
$\DS_5$ is a real, antisymmetric operator whose non-zero eigenvalues
are purely imaginary and occur in pairs. Any flow of these eigenvalues
under the smooth deformation of the five dimensional
gauge field will change the 
number of zero modes by a multiple of two, 
leaving the sign of the determinant intact. 
Now note that if two four dimensional gauge 
configurations are smooth deformations of each other, the five
dimensional configuration consisting of the interpolation 
between the two is itself smoothly deformable to a five dimensional
configuration which is just the constant (in $x_5$) interpolation
one obtains by extending one of the four dimensional configurations
into the fifth dimension. The latter will of course exhibit no
level crossing, so the former must exhibit only an even number
of crossings. Thus the signs of the determinants of two
four dimensional gauge configurations must be the same if they
are smoothly related.

Using the above result, the sign of the determinant for
arbitrary (four dimensional) $A_\mu (x)$ can be determined by 
using the index theorem
on a suitable vacuum configuration (\ref{pg}) which is smoothly connected to
$A_\mu (x)$.
Due to the factor of $2 C(R)$ from the index theorem, 
an integer--valued Casimir then
guarantees that the eigenvalues of $\DS_{\rm \, Dirac}$ 
used to define the Weyl
determinant only cross zero in even multiples, preserving the
sign of the Weyl determinant.

It remains to examine whether an arbitrary (four dimensional)
$A_\mu (x)$ can be 
smoothly connected
to some ``nearest'' vacuum configuration (\ref{pg}). 
In order to do this, 
we first carefully
examine the boundary conditions imposed on our field configurations. 
In order to
obtain a $\pi_4$ classification of field configurations in five 
dimensions, we require that the field approach a pure gauge on the
surface at infinity:
\beq
\label{bc}
A_\mu ( \vert x \vert \rightarrow \infty ) ~\rightarrow~ 
i U^{\dagger} \partial_{\mu} U (x)~~~.
\eeq
The functions $U(x)$ map $S^4 \rightarrow G$ and are classified by $\pi_4 (G)$,
allowing the application of the index theorem to $\DS_5$. If we consider 
the five
dimensional space to be $R^4 \times R$, then each slice at fixed $x_5$ obeys
boundary conditions like (\ref{bc}), but applied to the surface of $R^4$. This
allows a classification of each four dimensional configuration by $\pi_3 (G)$.
First consider the zero winding number sector. Here the vacuum is given by some
gauge function $U(x)$ which is smooth on all of $R^4$. By a smooth gauge 
transformation we can take $U(x) = U_0$ constant everywhere, so the vacuum 
is simply
$A_\mu = 0$. The boundary condition on a field configuration in this 
gauge is that
$A_\mu ( \vert x \vert \rightarrow \infty ) = 0$. A smooth interpolation which
relates $A_\mu$ to the vacuum is simply $A^t_\mu = t A_\mu$.

In the sectors with non-zero winding number under $\pi_3 (G)$, 
(ie non-zero
instanton number), there is a potential 
problem because extending the gauge function $U(x)$
from the surface at infinity ($S^3$) into the interior of $R^4$ cannot be done
without a singularity. This means that the ``nearest'' vacuum to a 
topologically
non-trivial configuration is itself singular, and there is the 
possibility that 
something discontinous can happen during the interpolation. However,
this possibility can be excluded because the interpolation from $A_\mu$ to
the vacuum is related by a large gauge transform to an interpolation in the
zero-winding sector which is smooth. Since the eigenvalues themselves are
gauge invariant, the interpolations in all sectors are smooth.

Actually, the zero-winding sector of configurations is sufficient to deduce the
properties of a model with zero theta angle. 
($\theta$ must be zero in any case to
preserve positivity of the functional measure.) 
This is because, in the infinite volume
limit, the only remnant of
the boundary conditions placed on the system is 
$\theta \int  F \tilde{F}$. 
When investigating the $\theta = 0$ theory, we are therefore allowed
to take any boundary conditions. In particular, we can define the theory in the
zero-winding sector without changing any of the physics.

The fourth homotopy group $\pi_4 ( G )$ 
is non-zero for $SU(2),~ O \, (N < 6)$ and $Sp(N)$ (any $N$).
These are the only cases in which the ambiguity can arise (although
this is far from clear {\it a priori}!).
For $SU(2)$, the case most likely to be of interest in lattice simulations
\cite{Montvay}, $\pi_4 ( SU(2) ) = {\bf Z}_2$.
In this case the Casimir of the adjoint representation is
$2$ ($C_{\rm adj} (~SU(N)~)  = N$), so  the sign of the Weyl determinant never 
fluctuates. In the other cases, we have
$C_{\rm adj} (~SO (2N+1)~)  = 4N - 2$,
$C_{\rm adj} (~SO (2N)~)  = 4N - 4$ and 
$C_{\rm adj} (~Sp (N)~)  = N + 1$,
so the determinants in these theories behave similarly.

We conclude by noting that our analysis is also of use in the study
of chiral gauge theories. In some recent proposals for the lattice realizations
of such theories \cite{chiralme,chiral}, the determinant is again constructed
from the product of half of the Dirac eigenvalues, with additional phase
information residing in functional Jacobian factors that result
from fermionic integration. 
Our analysis can be
used to determine whether there are sign fluctuations beyond those 
coming from the Jacobian factors. We see that unless the model 
is afflicted with a global anomaly (and hence inconsistent)
there are no such fluctuations.

\newpage
\begin{flushleft} {\Large\bf Acknowledgments} \end{flushleft}

\noindent 
The author would like to thank Nick Evans, Myckola Schwetz 
and Raman Sundrum
for useful discussions and comments.
This work was supported in part under DOE contract DE-AC02-ERU3075.


\end{document}